\documentclass[epj]{svjour}
\usepackage{graphics}
\begin{document}
%
%\title{Hyperon production in Ar+KCl collisions at 1.76A GeV}
\title{Hyperon production in Ar+KCl collisions at 1.76A GeV}
%\subtitle{}
\author{G.~Agakishiev$^{6}$, A.~Balanda$^{3,\dag}$, B.~Bannier$^{5}$, R.~Bassini$^{9}$,
D.~Belver$^{16}$, A.~Belyaev$^{6}$, A.~Blanco$^{2}$, M.~B\"{o}hmer$^{12}$, J.~L.~Boyard$^{14}$,
P.~Cabanelas$^{16}$
% ...
% etc
}                     % Do not remove
%
%\offprints{Lorenz@Physik.uni-frankfurt.de or aschmah@lbl.gov}          % Insert a name or remove this line
%
\institute{(HADES collaboration) \\
\mbox{$^{1}$Istituto Nazionale di Fisica Nucleare - Laboratori Nazionali del Sud, 95125~Catania, Italy}\\
\mbox{$^{2}$LIP-Laborat\'{o}rio de Instrumenta\c{c}\~{a}o e F\'{\i}sica Experimental de Part\'{\i}culas , 3004-516~Coimbra, Portugal}\\
\mbox{$^{3}$Smoluchowski Institute of Physics, Jagiellonian University of Cracow, 30-059~Krak\'{o}w, Poland}\\
\mbox{$^{4}$GSI Helmholtzzentrum f\"{u}r Schwerionenforschung GmbH, 64291~Darmstadt, Germany}\\
\mbox{$^{5}$Institut f\"{u}r Strahlenphysik, Forschungszentrum Dresden-Rossendorf, 01314~Dresden, Germany}\\
\mbox{$^{6}$Joint Institute of Nuclear Research, 141980~Dubna, Russia}\\
\mbox{$^{7}$Institut f\"{u}r Kernphysik, Goethe-Universit\"{a}t, 60438 ~Frankfurt, Germany}\\
\mbox{$^{8}$II.Physikalisches Institut, Justus Liebig Universit\"{a}t Giessen, 35392~Giessen, Germany}\\
\mbox{$^{9}$Istituto Nazionale di Fisica Nucleare, Sezione di Milano, 20133~Milano, Italy}\\
\mbox{$^{10}$Institute for Nuclear Research, Russian Academy of Science, 117312~Moscow, Russia}\\
\mbox{$^{11}$Excellence Cluster 'Origin and Structure of the Universe' , 85478~Munich, Germany}\\
\mbox{$^{12}$Physik Department E12, Technische Universit\"{a}t M\"{u}nchen, 85748~M\"{u}nchen, Germany}\\
\mbox{$^{13}$Department of Physics, University of Cyprus, 1678~Nicosia, Cyprus}\\
\mbox{$^{14}$Institut de Physique Nucl\'{e}aire (UMR 8608), CNRS/IN2P3 - Universit\'{e} Paris Sud, F-91406~Orsay Cedex, France}\\
\mbox{$^{15}$Nuclear Physics Institute, Academy of Sciences of Czech Republic, 25068~Rez, Czech Republic}\\
\mbox{$^{16}$Departamento de F\'{\i}sica de Part\'{\i}culas, Univ. de Santiago de Compostela, 15706~Santiago de Compostela, Spain}\\
\mbox{$^{17}$Instituto de F\'{\i}sica Corpuscular, Universidad de Valencia-CSIC, 46971~Valencia, Spain}\\
\mbox{$^{A}$ also at ISEC Coimbra, ~Coimbra, Portugal}\\
\mbox{$^{B}$ also at Technische Universit\"{a}t Dresden, 01062~Dresden, Germany}\\
\mbox{$^{C}$ also at Dipartimento di Fisica, Universit\`{a} di Milano, 20133~Milano, Italy}\\
\mbox{$^{D}$ also at Dipartimento di Fisica Generale, Universita' di Torino, 10125~Torino, Italy}\\
\mbox{$^{E}$ also at Joint Institute of Nuclear Research, 141980~Dubna, Russia}\\
\mbox{$^{F}$ also at Lawrence Berkeley National Lab, Berkeley California 94720, United States}\\
\\
\mbox{$^{\ast}$ corresponding author: Lorenz@Physik.uni-frankfurt.de, aschmah@lbl.gov}
\\
\mbox{$^{\dag}$ deceased}
}
\date{Received: 24.12.2010 / Revised version: date}
\abstract{
We present transverse momentum spectra, rapidity distribution and multiplicity of $\Lambda$-hyperons measured with the HADES spectrometer in the reaction Ar(1.76A GeV)+KCl. The yield of $\Xi^{-}$ is calculated from our previously reported $\Xi^{-}/(\Lambda+\Sigma^{0})$ ratio and compared to other strange particle multiplicities. Employing a strangeness balance equation the multiplicities of the yet unmeasured $\Sigma^{\pm}$ hyperons can be estimated.
Finally a statistical hadronization model is used to fit the yields of $\pi^-$, $K^+$, $K^0_s$, $K^-$, $\phi$, $\Lambda$ and $\Xi^-$. The resulting chemical freeze-out temperature of $T=(76\pm2)$ MeV is compared to the measured slope parameters obtained from fits to the transverse mass distributions of the different particles.}

%\authorrunning

%\titlerunning
\authorrunning{The HADES collaboration (G.~Agakishiev {\it et al.})}
\titlerunning{Hyperon production in Ar+KCl collisions at 1.76A GeV}
\maketitle

\section{Introduction}
\label{intro}
Strange hadrons are particularly suitable probes of the high density phase of nuclear matter produced in few GeV heavy ion collisions.  For instance, from systematic investigations of subthreshold K$^+$ production tight constraints could be put on the nuclear equation of state at matter densities of 2-3 $\rho_0$ \cite{Sturm01,Fuchs01,Hartnack02,Hartnack:2005tr}. Furthermore, kaon phase space distributions and flow patterns are considered to be sensitive to the in-medium kaon potential \cite{Schaffner97,Cassing97,Uhlig05}. On the other hand, due to strangeness conservation in the strong interaction, kaon production is intimately linked to the concurrent production of hyperons. While strange particle production is well understood in elementary NN collisions, in heavy ion reactions multi-step processes involving mesons or baryon resonances open up many additional production channels, even below threshold. Thus, strangeness-exchange channels like $\pi \Lambda \rightarrow N K^-$ have been proposed to explain the observed $K^-$ yields \cite{Hartnack:2001zs,Foerster03,Schade:2009gg}, just as feeding through the $\phi \rightarrow K^+K^-$ decay has been \cite{Agakishiev:2009ar}.

Various aspects of strangeness production at SIS (Schwerionen Synchrotron at GSI Darmstadt) energies have been investigated by the FOPI and KaoS experiments (for reviews see \cite{Foerster07,Fuchs06}). Evidently, any in-depth understanding of strangeness production and propagation in heavy ion reactions requires information on all particles with open or hidden strangeness. The HADES collaboration has done a complete measurement in the system Ar+KCl at a bombarding energy of 1.76A GeV. Results on $K^0_s$ production have been published already in \cite{Agakishiev:2010zw}, on $K^+, K^-$ and $\phi$-meson production in \cite{Agakishiev:2009ar} and on the first observation at such a low beam energy of the double-strange $\Xi^-$ hyperon in \cite{Agakishiev:2009rr}. To complete the picture, hyperon production remains to be addressed and this is the purpose of the present paper. We report here on the results obtained with the HADES detector on $\Lambda$ production, from which, by application of strangeness conservation, we could estimate also the yield of the (not directly observed) $\Sigma$ hyperons. Furthermore, we compare our set of particle yields to the result of a statistical hadronization model and discuss the implications. Note that the FOPI collaboration performed a similar analysis of strangeness production in the system Ni+Ni at 1.93A GeV \cite{Merschmeyer:2007zz}.

In Section 2 of this paper we give first a brief overview of the HADES detector and relevant details of the Ar+KCl data taking and then proceed to describe the employed particle identification and $\Lambda$ reconstruction procedures. In section 3 we present spectra and production yields of the ${\Lambda}$ hyperons. In section 4 the $\Lambda$ result is used to extract the yield of the double-strange ${\Xi^-}$ from our previously published ${\Xi^-}/{\Lambda}$ ratio \cite{Agakishiev:2009rr}. With all experimental yields established, strangeness balance is applied to estimate the yield of the unobserved charged ${\Sigma}$ hyperons. We compare all yields obtained in Ar+KCl with respect to a statistical hadronization model and confront the fitted chemical freeze-out temperature with the measured slope parameters obtained from transverse mass distributions. Finally we summarize our findings in section 5.

\section{ Experimental setup}
\label{experiment}
HADES is a charged-particle detector consisting of a 6-coil toroidal magnet centered on the beam axis and six identical detection sections located
between the coils and covering polar angles between $18^{\circ}$ and $85^{\circ}$.  Each sector is equipped with a Ring-Imaging Cherenkov (RICH)
detector followed by Multi-wire Drift Chambers (MDCs), two in front of and two behind the magnetic field, as well as a scintillator hodoscope (TOF/TOFino). Lepton identification is provided mostly by the RICH and supplemented at low polar angles with Pre-SHOWER chambers, mounted at the back of the apparatus. Hadron identification, however, is based on the time-of-flight and on the energy-loss information from TOF/TOFino, as well as from the MDC tracking chambers. A detailed description of HADES is given in \cite{Agakishiev:2009am}.

\begin{figure}
\begin{center}
\resizebox{5cm}{!}{%
  \includegraphics{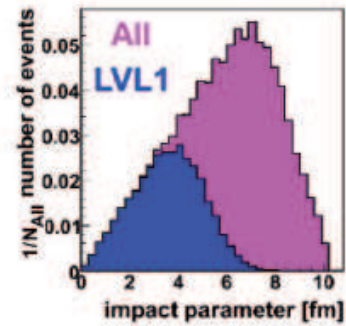}
% epja impact_parameter_herb.eps!!
}
\end{center}
%\vspace{5cm}       % Give the correct figure height in cm
\caption{Impact parameter distributions of all and LVL1 selected Ar+KCl reactions obtained from the UrQMD transport code \cite{UrQMD}.}
\label{fig_impact}       % Give a unique label
\end{figure}

An argon beam of $\sim 10^6$ particles/s was incident with a beam energy of 1.76A GeV on a four-fold segmented KCl target with a total thickness corresponding to $3.3$ $\%$ interaction probability. A fast diamond start detector located upstream of the target was intercepting the beam and was used to determine the time-zero information. The data readout was started by a first-level trigger (LVL1) requiring a charged-particle multiplicity, $MUL \ge 16$, in the TOF/TOFino detectors. Based on a full GEANT simulation of the detector response to Ar+KCl events generated with the UrQMD transport model \cite{UrQMD}, we found that the event ensemble selected by this (LVL1) trigger condition has a mean number of participating nucleons ($\langle A_{part}\rangle)$ equal to $38.5\pm3.9$.  Figure \ref{fig_impact} illustrates the impact parameter distributions obtained from UrQMD calculations for two event selections: all inelastic events and according to the experimental LVL1 trigger condition.

The particle identification was done by a velocity vs. momentum $\times$ polarity correlation, where the velocity was determined by the time-of-flight measurement in the TOF and TOFino scintillators with respect to the time-zero information and the tracked flight path.  If needed, additional particle discrimination was gained from the energy-loss (dE/dx) information in the MDC and scintillators.

%\section{Particle yields and strangeness conservation}
\section{$\Lambda$ hyperon yield and spectra}
\subsection{$\Lambda$ identification}
The particle identification of kaons, $\phi$, $\pi^-$ and $\Xi^-$ is described in \cite{Agakishiev:2009ar,Agakishiev:2009rr,Agakishiev:2010zw}. Here we add only those details specific to the reconstruction of $\Lambda$ hyperons in their decay channel $\Lambda \rightarrow p + \pi^-$ (B.R. =  63.9$\%$, c$\tau$=7.89 cm \cite{Amsler:2008zzb}). Note that at our beam energies the reconstructed $\Lambda$ yield contains also a contribution from decays of the slightly heavier $\Sigma^{0}$ baryon into a $\Lambda$ and a photon.

The decay products of the $\Lambda$ hyperons have been identified using the MDC dE/dx and time-of-flight information. The topology of the $\Lambda$ decay into p-$\pi^{-}$ pairs has been used to suppress the combinatorial background of uncorrelated pairs. Cuts on the distance between the primary event vertex and the decay vertex (d$_{V0}$), on the distances between the proton (d$_{p}$), respectively the $\pi^{-}$ (d$_{\pi^{-}}$) track and the primary vertex, on the distance of closest approach between the two tracks (d$_{dca}$) and on the distance of the reconstructed mother particle trajectory to the primary vertex (d$_{p\pi^{-}}$) were applied. Furthermore, a minimum opening angle ($\alpha_{p\pi^{-}}$) was required to guarantee a good decay vertex resolution. All selections used for the analysis are listed in Table \ref{tab_lambda_cuts}.
For an estimation of the systematic errors, the cut values have been varied within reasonable limits.  In total, 28 different cut combinations
were thus investigated, resulting in a total amount of reconstructed $\Lambda$ hyperons ranging from 36k to 191k. The full reconstruction chain with all corresponding efficiency corrections was applied for each of the cut variations. All systematic errors are based on these cut variations and represent the maximum/minimum deviation to the results obtained with the chosen cut values in Table \ref{tab_lambda_cuts}.

\begin{table*}
\caption{Topological conditions values chosen for the $\Lambda$ analysis (see text).}
\label{tab_lambda_cuts}       % Give a unique label
\begin{center}
\begin{tabular}{|l||c|c|c|c|c|c|}
    \hline
    Cut & d$_{V0}$  & d$_{p}$ &d$_{\pi^{-}}$  &  d$_{dca}$  &d$_{p\pi^{-}}$  &     $\alpha_{p\pi^{-}}$  \\
    \hline
    Value  & $>$70mm&$>$4.0mm & $>$d$_{p}$ & $<$10mm& $<$10mm &  $>$14$^{\circ}$ \\
    \hline
\end{tabular}
\end{center}
% Or use
%\vspace*{5cm}  % with the correct table height
\end{table*}

\begin{figure}
\begin{center}
\resizebox{8cm}{!}{%
  \includegraphics{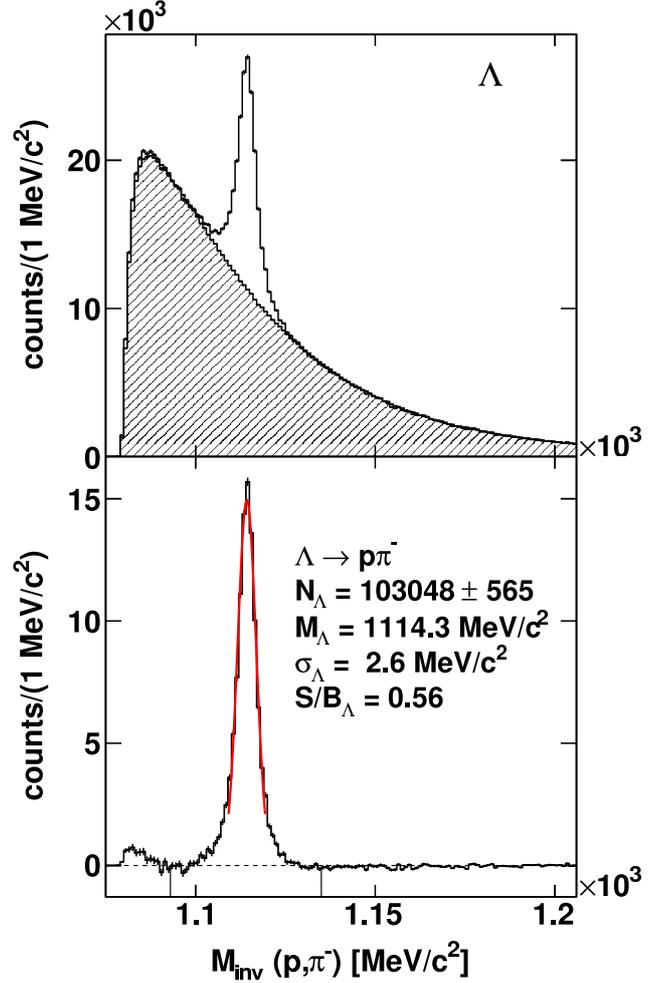}
}
\end{center}
%\vspace{5cm}       % Give the correct figure height in cm
\caption{Top: Invariant mass of all identified proton and $\pi^{-}$
     pairs after several cuts on the topology of the $\Lambda$ decay
     kinematics were applied (see text for details).  The grey shaded
     histogram shows the mixed-event combinatorial background, normalized to
     the signal spectrum between 1080-1100 and 1130-1150 MeV/c$^{2}$.
     Bottom: $\Lambda$ signal after background subtraction; the solid red line
     shows a Gaussian fit to the signal.}
\label{fig_lambda_mass}       % Give a unique label
\end{figure}

Figure \ref{fig_lambda_mass} shows the invariant-mass spectrum of all proton-$\pi^{-}$ pairs which passed the cuts listed in Table \ref{tab_lambda_cuts}.
An event-mixing technique has been used to model the combinatorial background of uncorrelated pairs. Displayed in Fig. \ref{fig_lambda_mass} as a grey shaded histogram, the mixed event background was normalized to the data on the left and right side of the $\Lambda$ peak. In total, for the optimal cut selection listed in Table \ref{tab_lambda_cuts},
about 100000 $\Lambda$ hyperons were reconstructed, with a mean signal-to-background ratio of 0.56. From a Gaussian fit to the peak, the pole mass is determined to be 1114.3 MeV/c$^{2}$, i.e. about 1.4 MeV/c$^{2}$ away from its listed value \cite{Amsler:2008zzb}. We attribute this small difference to residual deficiencies of our track reconstruction and detector alignment.

\subsection{$\Lambda$ spectra}

\begin{figure}
\resizebox{7cm}{!}{%
  \includegraphics{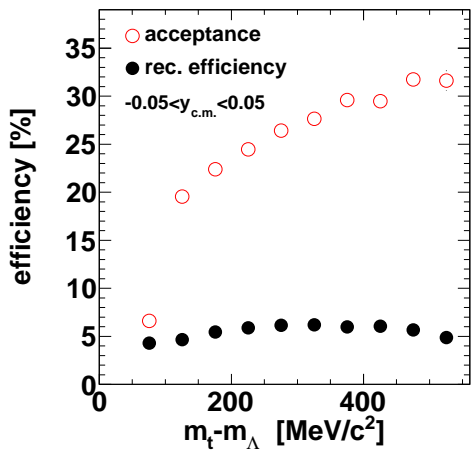}
}
\caption{(Color online) Acceptance and reconstruction efficiency at mid-rapidity for $\Lambda$-hyperons as a function of the reduced mass. The plotted acceptance (red open circles) includes already the branching ratio of 0.639 of $\Lambda \rightarrow p+\pi^{-}$. The reconstruction efficiency (black filled circles) consists of the single track reconstruction efficiencies of the daughter particles and the efficiencies of the topological cuts applied to improve the signal-to-background ratio (see text for details).}
\label{fig_eff}       % Give a unique label
\end{figure}

For further kinematical studies the $\Lambda$ signal has been determined in nine rapidity bins, ranging from $-0.75 < y_{c.m.} < +0.15$ in steps of 0.1, and up to ten transverse mass bins in steps of 50 MeV/c$^{2}$. The background subtracted signal yields were corrected for acceptance and reconstruction efficiency using a full GEANT simulation of the detector system described in \cite{Agakishiev:2010zw} and a track-embedding method. The geometrical acceptance, which also includes the branching ratio of $\Lambda \rightarrow p+\pi^{-}$ of 0.639, shows a smooth behaviour as a function of the transverse mass and varies for most of the bins between 13\% and 34\%.  It is defined by the requirement that both daughter particles have hits in all MDC planes. The $\Lambda$ reconstruction efficiency is composed of the single track reconstruction and particle identification efficiencies ($\approx$80\% per track) and the cuts on the $\Lambda$-hyperon decay topology. The latter one clearly dominates the reconstruction efficiency which has values of 3\% to 10\%. The dominant topology cut is the one applied on the distance between the primary vertex and the $\Lambda$-decay vertex. It is in the order of the $\Lambda$-hyperon mean decay length (see Table \ref{tab_lambda_cuts}). Acceptance and reconstruction efficiency are plotted for the mid-rapidity region in Fig. \ref{fig_eff}.

\begin{figure}
\resizebox{8.4cm}{!}{%
  \includegraphics{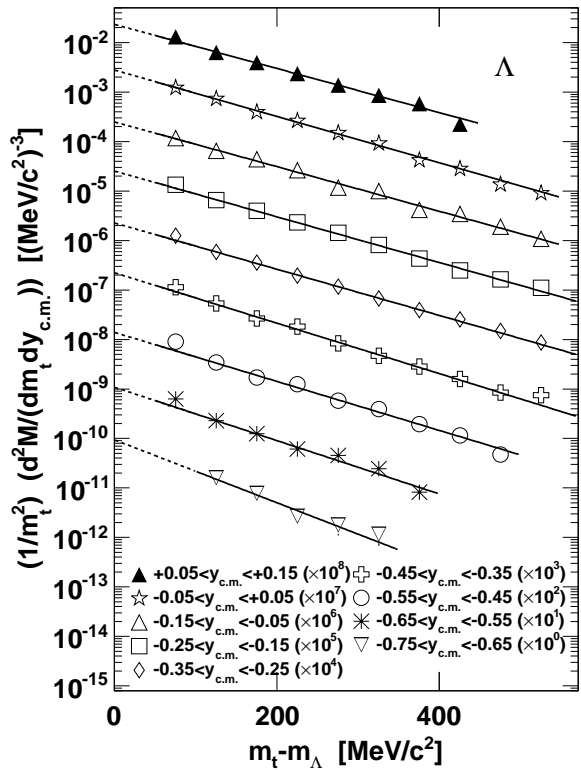}
}
%\vspace{5cm}       % Give the correct figure height in cm
\caption{Reduced ($m_{t}-m_{\Lambda}$) transverse mass spectra for
     different rapidity selections.  For better legibility, the spectra are
     scaled as indicated in the legend.  The solid lines are fits with Eq.
     \ref{bolz_eqn} to the data.}
\label{fig_lambda_mt}       % Give a unique label
\end{figure}

The acceptance- and efficiency-corrected transverse-mass spectra of $\Lambda$ for the various rapidity bins are presented in Fig. \ref{fig_lambda_mt}. Shown is the number of counts per LVL1 trigger, per transverse mass and per rapidity bin, divided by $m_t^2$. This representation is chosen in order to easily apply Boltzmann fits to the resulting distributions, according to
\begin{equation}
\label{bolz_eqn}
\frac{1}{m_{t}^{2}} \frac{d^2M}{dm_{t}dy_{c.m.}} = C(y_{c.m.}) \,
\exp \left( -\frac{(m_t-m_0)c^2}{T_B(y_{c.m.})}  \right) .
\end{equation}

\begin{figure}
\resizebox{8cm}{!}{%
  \includegraphics{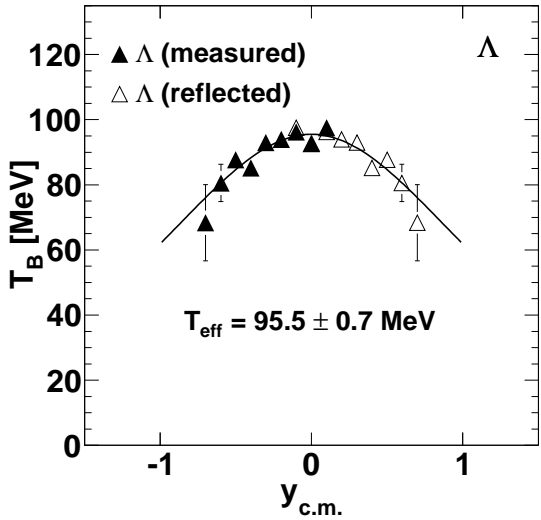}
}
\caption{Inverse-slope parameters from fits with Eq. (\ref{bolz_eqn}) to
     the $\Lambda$ hyperon transverse mass spectra as a function of rapidity.
     The solid line is a fit with Eq. (\ref{TB_func}) to the data points.
     The effective temperature $T_{eff}$ is the function value at mid rapidity.}
\label{fig_lambda_Tb}       % Give a unique label
\end{figure}

The solid lines in Fig. \ref{fig_lambda_mt} show the results of Boltzmann fits, where $T_B(y_{c.m.})$ represents the inverse slope of each distribution. The resulting $T_B(y_{c.m.})$ values are then plotted in Fig. \ref{fig_lambda_Tb} as a function of the center-of-mass (cm) rapidity ($y_{c.m.}=y-y(cm)$, where $y(cm)=0.858$ for symmetric collisions at 1.76A GeV). The full symbols display the measured data, whereas the open ones are the data reflected at c.m. rapidity. The error bars represent the statistical errors. Assuming a thermal source, these temperatures are expected to follow the relation
\begin{equation}
\label{TB_func}
T_B(y_{c.m.}) = \frac{T_{eff}}  {\cosh(y_{c.m.})},
\end{equation}
yielding an effective temperature of $T_{eff} =$ (95.5 $\pm 0.7 \textrm{(stat.)}$ $+2.2 \textrm{(syst.)})$ MeV. Here the systematic error corresponds to the variation of the cut values described above.

\begin{figure}
\resizebox{8cm}{!}{%
  \includegraphics{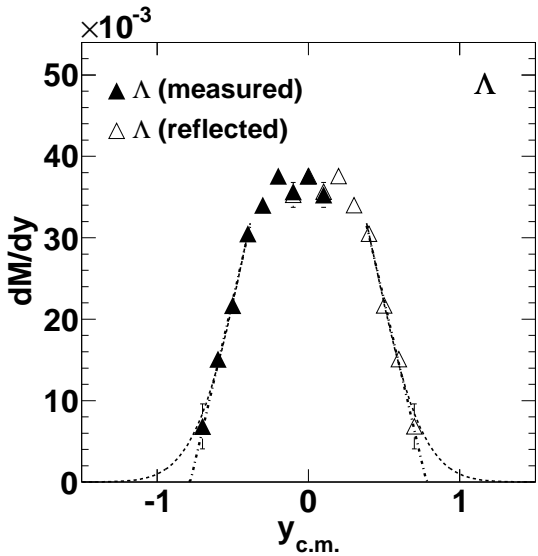}
}
\caption{Rapidity distribution of $\Lambda$ hyperons.  The closed
     symbols refer to measured data points calculated from the
     transverse mass spectra, whereas the open symbols show the data points
     reflected about center-of-mass rapidity.  For the extrapolation to
     unmeasured rapidity values a linear and a Gaussian function were
     fitted to the first four data points (see text).}
\label{fig_lambda_dNdy}       % Give a unique label
\end{figure}

For each rapidity bin the transverse mass spectrum was integrated in the following way: the yields in the covered bins were added and the fits were used to extrapolate into the unmeasured kinematic regions. The fits were integrated from 0 to the lower bin edge of the first measured point and from the upper bin edge of the last measured point to infinity.
The fraction of the extrapolated yield in the transverse mass spectra to the total yield is 36-43\% for the rapidity bins in the range $-0.65<y_{c.m.}<0.15$, and 65\% for the rapidity interval ($-0.75<y_{c.m.}<-0.65$). The results are shown in Fig.  \ref{fig_lambda_dNdy}, where the obtained
rapidity density distribution is displayed. The full triangles show the values calculated by the integration of the transverse mass spectra,
while the open triangles represent points reflected with respect to the center-of-mass rapidity.

For the determination of the total $\Lambda$ multiplicity per LVL1 event, the measured spectra were integrated. The extrapolation into the unmeasured region was done by fitting either a gaussian or a linear function to the first four data points, as shown in Fig. \ref{fig_lambda_dNdy}. The mean value of these two different extrapolations is used for the total yield. The fraction of the Gaussian extrapolation to the total yield is about 4.2\%, whereas the fraction of the linear extrapolation is negligible. The inclusive total $\Lambda$ multiplicity per LVL1 event was found to be
$(4.09 \pm 0.1 \textrm{(stat.)} \pm 0.17 \textrm{(extr.)} ^{+0.17}_{-0.37} \textrm{(syst.)}) \times 10^{-2}$, where the second error refers to the extrapolation uncertainty in $m_{t}$ and the third one to the systematic error obtained from the cut variations. % as well as the feed down from weak decays.
A detailed description of the $\Lambda$ analysis can be found in \cite{PhD_Schmah}.

\section{Discussion}
\subsection{Particle yields}
Knowing the yield of the $\Lambda + \Sigma^{0}$ hyperons and the $\Xi^{-}$/$(\Lambda + \Sigma^{0})$ ratio yields the production rate of the $\Xi^{-}$ to $(2.3\pm0.9) \times 10^{-4}$, adding statistical and systematic errors quadratically. Note that this value is of the same order of magnitude as the yield of the $\phi$ meson \cite{Agakishiev:2009ar}.

Table \ref{mult_table} summarizes all particle yields extrapolated to full phase space as well as the corresponding inverse slope parameters from fits to the particle $m_{t}$ spectra;  results on $K^+, K^-, K^0_s$, and $\phi$ are taken from \cite{Agakishiev:2009ar,Agakishiev:2010zw}.

\begin{table*}
\caption{Multiplicities (i.e. yield/LVL1 event) and effective temperatures
  of particles produced in Ar+KCl reactions at 1.76A GeV. The error
  on the $\Sigma$ and $\Xi^{-}$ yield is the quadratically added statistical and
  systematic error.}
\label{mult_table}       % Give a unique label

\begin{center}
\begin{tabular}{|c|c|c|c|}
    \hline
    Particle & Multiplicity & $T_{eff}$ [MeV] & Reference\\
    \hline
    \hline
    $\pi^-$ & $3.9 \pm 0.1 \pm 0.1$  & $82.4 \pm 0.1 ^{+9.1}_{-4.6}$& \cite{Agakishiev:2010zw} \\
   \hline
    %$p$ & $22.11$  & $144.0$& this work \\
    %\hline
    $\Lambda+\Sigma^0$ & $(4.09 \pm 0.1  \pm 0.17 ^{+0.17}_{-0.37}) \times 10^{-2}$ & $95.5 \pm 0.7  +2.2$ & this work \\
    \hline
    $K^+$  & $(2.8 \pm 0.2 \pm 0.1 \pm 0.1) \times 10^{-2}$ & $89 \pm 1 \pm 2$ & \cite{Agakishiev:2009ar} \\
    \hline
    $K^{0}_{S}$ & $(1.15 \pm 0.05 \pm 0.09) \times 10^{-2}$ & $92\pm 2$ & \cite{Agakishiev:2010zw} \\
    \hline
    $K^-$ & $(7.1 \pm 1.5 \pm 0.3 \pm 0.1) \times 10^{-4}$  & $69 \pm 2 \pm 4$& \cite{Agakishiev:2009ar} \\
    \hline
    %$\omega$ & $(6.7 \pm 2.7) \times 10^{-3}$  & & cite \\
    %\hline
    $\phi$ & $(2.6 \pm 0.7 \pm 0.1 -0.3) \times 10^{-4}$  & $84 \pm 8$& \cite{Agakishiev:2009ar} \\
    \hline
    $\Xi^{-}$ & $(2.3 \pm 0.9) \times 10^{-4}$  & - & \cite{Agakishiev:2009rr} \\
    \hline
    \hline
     $\Sigma^{+}+\Sigma^{-} $ & $(0.75 \pm 0.65) \times 10^{-2}$  & - & estimated via strangeness balance \\
    \hline
    \end{tabular}
\end{center}
\end{table*}

\subsection{Strangeness balance}
The strong interaction conserves strangeness, i.e. the numbers of $s$ and $\overline{s}$ quarks produced in a heavy ion reaction must be equal.  As those quarks are ultimately bound in hadrons the multiplicities of strange particles fulfill a balance equation which can be written at SIS energies as:
\begin{equation}
K^+ + K^0  = \Sigma^{0\pm} +\Lambda + K^- +\bar{K^0} + 2\Xi^{0,-}
\label{str_eqn1}
\end{equation}
where, for simplicity, the symbols denoting the particles stand for their respective yields at the time of production. Note that this equation takes care of the strong decay of heavier strange resonances via the counting of their decay products, namely kaons and $\Lambda$s. As mentioned earlier, the $\Sigma^0$ can not be separated from the $\Lambda$, thus this contribution is to be counted explicitly together with $\Lambda$s. Analogously, according to our analysis procedure, most $\Xi^{-,0}$ decay products feed the $\Lambda$ channel and are counted as $\Lambda$s, therefore the factor in Eq. (\ref{str_eqn1}) in front of the $\Xi^{-,0}$ is two instead of four. (Note that anyhow the $\Xi^{-,0}$ contributions are small.) In case of the neutral kaons, we measure in fact the yield of the $K^0_{s}$, which obeys the equality $K^0_{s} = (K^0+\bar{K^0})$/2. Assuming isospin symmetry the yield of the $\bar{K^0}$ should be contributing here at the same order as the $K^-$ yield.
Eq. (\ref{str_eqn1}) can then be rewritten using the measured yields. Hence the unobserved $\Sigma^{\pm}$ hyperon yield can be estimated as:
\begin{equation}
\Sigma^+ + \Sigma^- = K^+ + 2K^0_{s}  - (\Sigma^0 + \Lambda ) - 2\Xi^{-} - 3K^-
\label{str_eqn}
\end{equation}
Still heavier multi-strange particles, e.g. $\Omega$ hyperons, have significantly higher production thresholds and should not contribute sizeably at SIS energies. From Eq. (\ref{str_eqn}) a total multiplicity of charged $\Sigma$ hyperons of $(7.5 \pm 6.5)\times 10^{-3}$ is deduced when using the values of the multiplicities listed in Table \ref{mult_table}. The error is the quadratic sum of the statistical and systematic error of the different yields.
If one assumes isospin symmetry for the $\Sigma^{\pm, 0}$ yields one can subtract the $\Sigma^{0}$ contribution from the $\Lambda$ yield and finds
the ratio $\Sigma^{\pm, 0}/\Lambda$ is $0.3\pm0.26$ although the difference in mass is only $10\%$.
 %that the difference between the $\Lambda$ and $\Sigma=\Sigma^{0}+\Sigma^{+}+\Sigma^{-}$ yield is around $70\%$ although the difference in mass is only $10\%$. %Note that the
%in yield is nearly $70\%$, even after correcting the $\Lambda$ yield for its $\Sigma^0$
%contribution (assuming isospin symmetry for the $\Sigma^{\pm, 0}$ yields).
%This might be related to a threshold effect, as the $\Lambda$ is produced here slightly above, while the $\Sigma$ is produced slightly below their respective free NN thresholds.

%Recent studies of $\Sigma$ production in pp collisions at small excess energies \cite{Budzanowski:2010df} find that the charged $\Sigma$ production is in accordance to phase space.
The only other published multiplicity of charged $\Sigma$ hyperons in heavy ion collisions, based on a similar analysis of strangeness yields measured with the FOPI detector at GSI in Ni+Ni reactions at 1.93A GeV, is $(7\pm8+32-17)\times 10^{-3}$ \cite{Merschmeyer:2007zz}. Differences with respect to our analysis are: (1) a higher beam energy (1.93 vs. 1.76A GeV), (2) a larger reaction system (58+58 vs. 40+37), and (3) a different centrality selection ($ \langle A_{part} \rangle$ = 71 vs. 38.5) resulting in a higher $\Lambda$ yield ($0.137\pm0.005+0.007-0.008$) and inverse slope parameter ($119\pm1+9-7$).  In view of the higher bombarding energy and larger system size, one would expect a larger charged $\Sigma$ contribution. The resulting $\Sigma^{\pm, 0}/\Lambda$ ratio is $0.08\pm0.09+0.33-0.18$. Unfortunately, the large uncertainties prevent us to draw a firm conclusion on the behavior of the respective cross sections with energy.
%It could point to a rather complex behavior of the respective cross sections with energy, however one has to keep the large uncertainties in mind.  %{\em Was sagt denn Thermus dazu ?}
%From the strangeness balance one can also calculate the charged-to-neutral
%hyperon production ratio
%\begin{equation}
%\frac{\Sigma^+ + \Sigma^-}{\Lambda + \Sigma^0} = 0.18 \pm 0.14
%\end{equation}
%which clearly favors neutrals.

%The result may be compared with the result from recent studies of $\Sigma$ production in \pp collisions at small excess energy \cite{Budzanowski:2010df} which found that the charged $\Sigma$ production is in accordance to phase space. %It is hence natural to turn to a statistical description of the observed yields, which we do in the next section.

\subsection{Comparison with statistical hadronization}

Statistical hadronization models (SHM) have been successful in fitting particle yields or yield ratios from relativistic and ultrarelativistic heavy ion collisions \cite{Averbeck:2000sn,Becattini:2003wp,BraunMunzinger:2003zd,Cleymans:2005xv}. With the help of SHM fits it has been possible to reconstruct systematically the chemical freeze-out line in the T -- $\mu_{b}$ plane of the nuclear phase diagram with $\mu_{b}$ being the baryochemical potential
(see e.g. \cite{Cleymans:2005xv,Andronic:2009gj}). However, while the various SHM approaches agree fairly well at high bombarding energies, discrepancies appear in the low-energy regime. % of the SIS and the Alternating Gradient Synchrotron (AGS), (see Fig. \ref{fig_TmuB}).
%{\em Kann man das quantifizieren?  Ref?}.
Indeed, at the lower energies it is not even clear, whether chemical equilibrium can be reached \cite{Koch:2003pj} and therefore the question arises whether a statistical treatment of particle production is meaningful. The situation is further complicated by the need for strangeness suppression, which is handled differently in the various SHM implementations. Furthermore, at SIS energies, only pions are produced abundantly. Heavier and especially strange particles are rare, and their yields were mostly poorly known. Hence in the past only few particle yields with small statistical errors were available as input to the fit procedure. In the following we fit eight particle yields obtained from our Ar+KCl run with a statistical hadronization model.

\begin{figure}
\resizebox{8cm}{!}{%
  \includegraphics{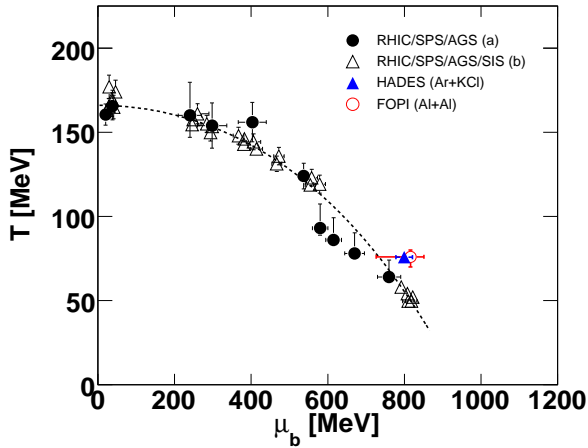}
}
\caption{(Color online) Chemical freeze-out points in the T --
    $\mu_{b}$ plane. The filled black circles (a) are taken from
    \cite{Andronic:2005yp}, the black open triangles (b) are
    from \cite{Cleymans:2005xv}. The red circle is taken from \cite{Lopez:2007ms}.
    The THERMUS fit to our Ar+KCl data is shown as blue triangle.
    The dashed line correponds to a fixed energy per nucleon of 1 GeV,
    calculated according to \cite{Cleymans:2005xv}.}
\label{fig_TmuB}       % Give a unique label
\end{figure}

We choose the freely available THERMUS code \cite{Wheaton:2004qb}, using the mixed canonical ensemble where strangeness is exactly conserved while all other quantum numbers are calculated grand canonically. We handle the strangeness suppression by introducing a strangeness correlation radius $R_{c}$ within which strangeness has to be exactly conserved; this is discussed in \cite{Kraus:2007hf}. We fit simultaneously all particle yields  listed in Table \ref{mult_table} except for the $\Sigma^{\pm}$, as well as the mean number of participants $\langle A_{part} \rangle$ and constrain the charge chemical potential $\mu_{Q}$ using the ratio of the baryon and charge numbers of the collision system. We find the chemical freeze-out at a temperature of $T_{chem}=(76\pm2)$ MeV and at a baryochemical potential of $\mu_{b}=(799\pm22)$ MeV. The strangeness correlation radius comes out as $R_{c}=(2.2\pm0.2)$ fm, which corresponds to about half the fitted radius $R=(4.1\pm0.5)$ fm of the whole fireball. %({\em Apart=39 $\rightarrow$ R=4.1 fm?}).
%The reduced $\chi^2$ value of this fit is 11.4.
The exclusion of the $\Xi^{-}$ from the fit changes the parameters only on the percent level, but the $\chi^2$/d.o.f. value of the fit improves from 13.9/4 to 7.8/3.
Fig. \ref{fig_TmuB} shows the resulting freeze-out point together with a compilation of similar points \cite{Andronic:2005yp,Cleymans:2005xv,Lopez:2007ms} in the T -- $\mu_{b}$ plane. Our result, as well as the FOPI result from the collision system Al+Al at 1.9A GeV, differ from the regularity of freeze-out points following the fixed energy per particle condition $\langle E \rangle / \langle N \rangle \approx 1$ GeV, which is one of the commonly proposed freeze-out criteria \cite{Cleymans:2005xv}. This might be due to the light collision systems, since small systems have the tendency to show higher freeze-out temperatures \cite{Cleymans:1999}.

\begin{figure}
\resizebox{8cm}{!}{%
  \includegraphics{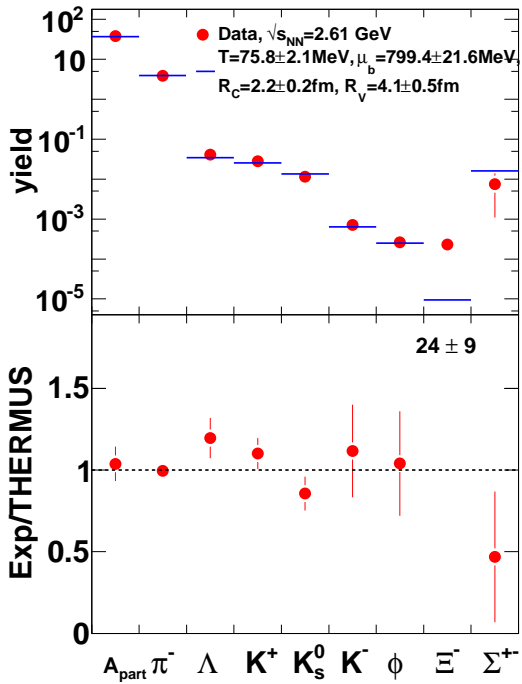}
}
\caption{(Color online) The upper plot shows the yields of secondary hadrons in Ar+KCl reactions (filled red
     circles) and the corresponding THERMUS fit (blue bars). The lower plot shows the ratio of the experimental value and the SHM value. For the $\Xi^{-}$ the ratio number is quoted instead of a point.}
\label{fig_Thermus_fit}       % Give a unique label
\end{figure}
A detailed comparison of the data with the statistical model fit is shown in the upper part of Fig. \ref{fig_Thermus_fit}, while the lower part of this figure depicts the ratio of data and fit. All particles except for the $\Xi^-$ are well described.
%except for small deviations in the $\Lambda$ yield, which,because of feeding, also affect the estimation of the charged $\Sigma$ fraction. This might be due to the threshold effect mentioned in the strangeness balance discussion above. Since the $\Lambda$ is produced slightly above its free NN-threshold while the $\Sigma$ is slightly below, the data might reflect the different thresholds stronger than the statistical calculation which populates the phase space in accordance to the temperature T and the baryochemical potential $\mu_{B}$.

A particularly interesting case is the $\phi$ meson. The $\phi$ is treated as a strangeness neutral object in the $R_{c}$ formalism and is therefore not suppressed at all. Its yield is well described by the SHM. This means that the $\phi$ yield is compatible with the assumption that it takes part in the equilibration of the hadrons. This is quite different from the situation at higher bombarding energies, where the $\phi$ requires indeed an effective strangeness between 1 and 2 to have the appropriate suppression in the SHM and to reproduce the data \cite{Kraus:2007hf}. For an understanding of $\phi$ production, one may have a look at the $\phi$/$K^-$ ratio which, according to the SHM with $R_{c}$, should rise at low beam energy. Such a behavior is indeed supported by our data, as already discussed in \cite{Agakishiev:2009ar}, and the ratio seems to approach the value seen in elementary NN reactions \cite{Maeda:2007cy}.
According to the strangeness suppression mechanism implemented in SHM, the double-strange $\Xi^-$ (S=2) should be suppressed strongly with respect to the $\phi$ with its hidden strangeness (S=0).  Nevertheless, our measured $\Xi^-$ yield is of the same magnitude as the one of the $\phi$, i.e. the data show no indication for any strangeness suppression. This is very surprising since the $\Xi^-$ yields observed above threshold at RHIC \cite{Adams:2006ke}, at SPS \cite{Antinori:2004ee} and even at AGS \cite{Chung:2003zr} are consistent with statistical model fits. In fact the same secondary pion-hyperon process $\pi + Y \rightarrow \phi + Y$, which was invoked by Kolomeitsev and Tomasik \cite{Kolomeitsev:2009yn} to explain the enhanced $\phi$ yield, can here be the origin of the high $\Xi$ production via the reaction $\pi + Y \rightarrow \Xi + K$.
To get a better understanding, we may have to move away from the SHM. One may calculate the probability for the production of two $s\overline{s}$ pairs in one collision.  Assuming that both pairs are independently created, their production probability $P_{2s\overline{s}}$ is given as the square of the single-pair production probability $P_{s\overline{s}}$.  Keeping associated production in mind, $P_{s\overline{s}}$ can be estimated as the combined multiplicity of all particles that carry a strange quark, respectively the combined multiplicity of all anti-strange particles, i.e.
$K^+ + K^0 + \phi$, yielding $P_{s\overline{s}} \simeq 0.05$ and hence $P_{2s\overline{s}} \simeq 0.0025$.  Considering that the observed $\Xi^-$ yield is in fact an order of magnitude smaller, we conclude that in 10\% of these events both $s$ quarks end up together in a $\Xi^-$, whereas from strangeness suppression in the SHM one obtains less than 1\%.
A different realization of the SHM using the strangeness canonical ensemble and $\gamma_{s}$ for additional strangeness suppression delivers comparable freeze-out parameters, with $T_{chem}=(76\pm5)$ MeV, $\mu_{b}=(791\pm33)$ MeV, $R=(4.1\pm0.9)$ fm and $\gamma_{s} =0.37\pm0.04$ but fails to reproduce the $\phi$ multiplicity by an order of magnitude due to its suppression with $\gamma_{s}^2$.

\begin{figure}
\resizebox{8cm}{!}{%
  \includegraphics{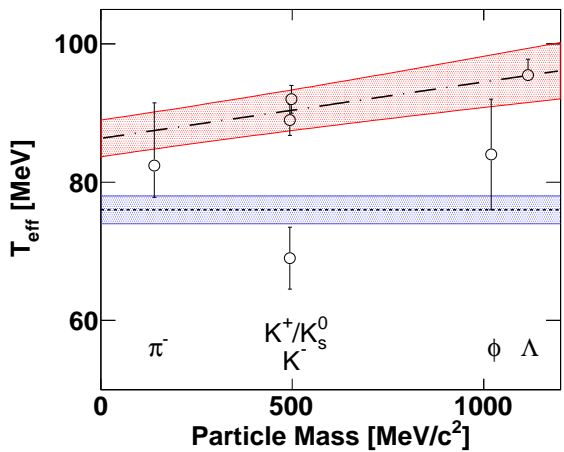}
}
\caption{(Color online) Effective temperature $T_{eff}$ of all measured particle
     species as a function of their mass. The horizontal line and error
     band show the chemical freeze-out temperature $T_{chem}$ from the THERMUS fit, whereas the dashed-dotted line and the red error band
     show a linear fit to the data points ($K^{-}$ are excluded, see text for details).}
\label{fig_slopes}       % Give a unique label
\end{figure}

\subsection{Chemical  vs.  kinetic freeze-out}
The temperature $T_{chem}$ obtained for the chemical freeze-out can be compared with the inverse-slope parameter $T_{eff}$ obtained from Boltzmann fits to the $m_{t}$ spectra of the different particle species. Apparently most of the inverse-slope parameters are higher than the chemical freeze-out temperature of the system.
%From Fig. \ref{fig_slopes} one sees that there is no strong difference between $T_{eff}$ and $T_{chem}$.
% with, however, one prominent outlier : the proton.  Such a large
%apparent proton temperature has also been seen by FOPI in Ni+Ni \cite{Merschmeyer:2007zz},
%as well as even much earlier in Bevalac Ne+NaF data \cite{Nagamiya:1981}.
A pure Boltzmann shape can be distorted by various effects, like collective motion or early vs. late particle decays. One example is apparent in the difference between the $K^+/K^0_s$ and $K^-$ slopes (see Fig. \ref{fig_slopes}). The much lower value of $T_{eff}$ of the $K^-$ has often been interpreted as due to its much later freeze-out time \cite{Foerster07} neglecting the admixture of soft $K^-$ stemming from $\phi$ decays. In \cite{Bormio} however it was shown that these soft $K^-$ indeed affect the shape of the spectra.

Effects of collective flow, on the other hand, should influence the transverse mass slope more, the higher the particle mass. From Fig. \ref{fig_slopes}, where the fitted temperatures are ordered by increasing particle mass, this seems not to be a strong effect as expected for a small collision system like Ar+KCl. %\footnote{Similar findings have been recently reported by the FOPI collaboration analyzing the transverse temperatures of hydrogen isotopes \cite{reisdorf}.}
However the inverse-slope parameter seems to be slightly decreasing with decreasing mass.

To take this effect into account a linear fit to the data points was applied (dashed-dotted line and red error band in Fig. \ref{fig_slopes}). The $K^{-}$ were excluded from the fit for the reasons discussed above. The fit is clearly dominated by the $\Lambda$ and $K^+/K^0_{s}$ data points with small relative errors compared to the $\phi$ and $\pi^{-}$ data points. Within errors, the extrapolated fit value at mass 0 is still above the chemical freeze-out temperature. Hence the presented data implies an inversion of the kinetic- and chemical freeze-out scenario, which cannot be the case for obvious reasons. This means either the statistical model approach for small reaction systems and small energies is not applicable, and/or the unique kinetic freeze-out for all particles with one radial flow velocity is a too naive assumption for this system. Future measurements of HADES in the reaction systems Au+Au and Ag+Ag will give more insight into the complex dynamics at low energies.

%Transport models on the other hand are able to describe kaon yields, as well as kinematical observables with good agreement since many years. While %only recent calculations are able to describe the $\phi$ yield \cite{Schade:2009gg}, the high $\Xi$ yield remains unexplained by transport up to now.
%As pointed out above,
%the proton makes exception, but then it is not a produced particle and the authors
%of \cite{Merschmeyer:2007zz} interpreted its behavior as a sign for incomplete
%stopping in medium-sized systems.

\section{Summary and conclusions}
We have presented phase space distributions of $\Lambda$ hyperons in Ar+KCl at 1.76A GeV measured with the HADES spectrometer at GSI.  Combining the
measured $\Lambda + \Sigma^0$ yield with our former data on strangeness production in this system we have estimated the yield of the double-strange $\Xi^{-}$ hyperon. We find that it is of the same order of magnitude as the one of the $\phi$ meson. The fraction of the unobserved charged $\Sigma^{\pm}$ hyperons could be constructed using strangeness conservation.

Applying a statistical model fit to these hadron yields, a fair agreement, except for the $\Xi^{-}$, in a strangeness-canonical approach is achieved. The $\phi$, however, is well reproduced without any suppression, in sharp contrast to the situation at higher energies, where a suppression is observed. %The mechanism responsible for the relatively large $\Xi^-$ production remains unclear and needs more investigation, both experimentally and theoretically.

The HADES collaboration gratefully acknowledges the support by BMBF grant 06MT9156, 06GI146I,06FY171 and 06DR9059D (Germany), by GSI (TMKrue 1012, GI /ME3, OF/STR), by Excellence Cluster Universe (Germany), by grants GA AS CR IAA100480803 and MSMT LC 07050 MSMT (Czech Republic), by grant KBN5P03B 140 20 (Poland), by INFN (Italy), by CNRS/IN2P3 (France), by grants MCYT FPA2000-2041-C02-02 and XUGA PGID FPA2009-12931 T02PXIC20605PN (Spain), by grant UCY-10.3.11.12 (Cyprus), by INTAS grant 06-1000012-8861 and EU contract RII3-CT-506078.
%We thank all funding agencies for spending money on this project. We love
%you all!


\begin{thebibliography}{9}
% Evidence for a soft EOS from kaon production in HIC:
\bibitem{Sturm01} C. Sturm {\it et al.} (KaoS Collaboration), Phys. Rev. Lett. \textbf{86}, 39 (2001).
% Probing the nuclear EOS by K+ production in HIC:
\bibitem{Fuchs01} C. Fuchs, A. Faessler, E. Zabrodin, Phys. Rev. Lett. \textbf{86}, 1974 (2001).
% Analysis of kaon production around the threshold
\bibitem{Hartnack02} C. Hartnack, J. Aichelin, J. Phys. G \textbf{28}, 1649 (2002).
%\cite{Hartnack:2005tr}
\bibitem{Hartnack:2005tr} C.~Hartnack, H.~Oeschler and J.~Aichelin, Phys.\ Rev.\ Lett.\ \textbf{96}, 012302 (2006).
\bibitem{Schaffner97} J. Schaffner-Bielich, J. Bondorf, A. Mishustin, Nucl. Phys. A \textbf{625}, 325 (1997).
\bibitem{Cassing97} W. Cassing, E.\,L. Bratkovskaya, U. Mosel, S. Teis, A. Sibirtsev,
Nucl . Phys. A \textbf{614} 415 (1997).
\bibitem{Uhlig05} F. Uhlig {\it et al.} (KaoS Collaboration), Phys. Rev. Lett. \textbf{95}, 12301 (2005).
% Different freeze out conditions of K* and K-
\bibitem{Hartnack:2001zs}
  C.~Hartnack, H.~Oeschler and J.~Aichelin,
  %``What determines the K- multiplicity at energies around 1-A-GeV to
  %2-A-GeV?,''
  Phys.\ Rev.\ Lett.\ \textbf{90}, 102302 (2003).
  %[arXiv:nucl-th/0109016].
  %%CITATION = PRLTA,90,102302;%%

\bibitem{Foerster03} A. F\"orster {\it et al.} (KaoS Collaboration), Phys. Rev. Lett. \textbf{91}, 152301 (2003).
%\cite{Agakishiev:2009ar}
%\cite{Agakishiev:2009ar}
%\cite{Hartnack:2001zs}
\bibitem{Schade:2009gg}
  H.~Schade, Gy.~Wolf and B.~K\"ampfer,
  %``Role of phi decays for K- yields in relativistic heavy-ion collisions,''
  Phys.\ Rev.\  C \textbf{81}, 034902 (2010).
  %[arXiv:0911.3762 [nucl-th]].
  %%CITATION = PHRVA,C81,034902;%%

\bibitem{Agakishiev:2009ar}
  G.~Agakishiev {\it et al.}  (HADES Collaboration),
  %``Phi decay: a relevant source for K- production at SIS energies?,''
  Phys.\ Rev.\  C \textbf{80}, 025209 (2009).
  %[arXiv:0902.3487 [nucl-ex]].
  %%CITATION = PHRVA,C80,025209;%%
\bibitem{Foerster07} A. F\"orster {\it et al.} (KaoS Collaboration), Phys. Rev. C 75, 024906 (2007).
%Kaon production in heavy ion reactions at intermediate energies
\bibitem{Fuchs06} C. Fuchs, Progr. Part. Nucl. Phys. \textbf{56}, 1 (2006).
\bibitem{Agakishiev:2010zw} G.~Agakishiev {\it et al.} (HADES Collaboration), Phys. Rev. C 82, 044907 (2010).
  %``In-Medium Effects on K$^0$ Mesons in Relativistic Heavy-Ion Collisions,''
%\cite{Agakishiev:2009rr}
\bibitem{Agakishiev:2009rr}
  G.~Agakishiev {\it et al.}  (HADES Collaboration),
  %``Deep sub-threshold $\Xi^-$ production in Ar+KCl reactions at 1.76A GeV,''
  Phys.\ Rev.\ Lett.\  \textbf{103}, 132301 (2009).
  %[arXiv:0907.3582 [nucl-ex]].
  %%CITATION = PRLTA,103,132301;%%
\bibitem{Merschmeyer:2007zz} M.~Merschmeyer {\it et al.} (FOPI Collaboration), Phys.\ Rev.\  C \textbf{76}, 024906 (2007).
  %``K0 and Lambda production in Ni+Ni collisions near threshold,''
\bibitem{Agakishiev:2009am}
  G.~Agakishiev {\it et al.}  (HADES Collaboration),
  %``The High-Acceptance Dielectron Spectrometer HADES,''
  Eur.\ Phys.\ J.\  A \textbf{41}, 243 (2009).
  %[arXiv:0902.3478 [nucl-ex]].
  %%CITATION = EPHJA,A41,243;%%
\bibitem{UrQMD} S.\,A. Bass {\it et al.}, Prog. Part. Nucl. Phys. \textbf{41} 225 (1998).
%\cite{Amsler:2008zzb}
\bibitem{Amsler:2008zzb} C.~Amsler {\it et al.}  (Particle Data Group), Phys.\ Lett.\  B \textbf{667}, 1 (2008).
\bibitem{PhD_Schmah} A. Schmah, PhD thesis, Technical University Darmstadt, Darmstadt (2008).
%Cosy
%\cite{Rozek:2006ct}
%\bibitem{Rozek:2006ct} T.~Rozek {\it et al.}, Phys.\ Lett.\  B {\bf 643}, 251 (2006) %[arXiv:nucl-ex/0607034].
  %``Threshold hyperon production in proton proton collisions at COSY-11,''
%\cite{Averbeck:2000sn}
%\cite{Budzanowski:2010df}
%\bibitem{Budzanowski:2010df}
%  A.~Budzanowski {\it et al.} (HIRES),
  %``Cross section of the $pp\to K^+\Sigma^+n$ reaction close to threshold,''
%  arXiv:1007.1542 [hep-ex] 2010.
  %%CITATION = ARXIV:1007.1542;%%


\bibitem{Averbeck:2000sn}
  R.~Averbeck, R.~Holzmann, V.~Metag and R.~S.~Simon,
  %``Neutral pions and eta mesons as probes of the hadronic fireball in  nucleus
  %nucleus collisions around 1-A-GeV,''
  Phys.\ Rev.\  C \textbf{67}, 024903 (2003).
  %[arXiv:nucl-ex/0012007].
  %%CITATION = PHRVA,C67,024903;%%
%\cite{Becattini:2003wp}
\bibitem{Becattini:2003wp}
  F.~Becattini, M.~Gazdzicki, A.~Keranen, J.~Manninen and R.~Stock,
  %``Study of chemical equilibrium in nucleus nucleus collisions at AGS and  SPS
  %energies,''
  Phys.\ Rev.\  C \textbf{69}, 024905 (2004).
  %[arXiv:hep-ph/0310049].
  %%CITATION = PHRVA,C69,024905;%%
%\cite{BraunMunzinger:2003zd}
\bibitem{BraunMunzinger:2003zd}
  P.~Braun-Munzinger, K.~Redlich and J.~Stachel,
  %``Particle production in heavy ion collisions,''
  arXiv:nucl-th/0304013 (2003).
  %%CITATION = NUCL-TH/0304013;%%
%\cite{Cleymans:2005xv}
\bibitem{Cleymans:2005xv}
  J.~Cleymans, H.~Oeschler, K.~Redlich and S.~Wheaton,
  %``Comparison of chemical freeze-out criteria in heavy-ion collisions,''
  Phys.\ Rev.\  C \textbf{73}, 034905 (2006).
  %[arXiv:hep-ph/0511094].
  %%CITATION = PHRVA,C73,034905;%%

\bibitem{Cleymans:1999} J.~Cleymans {\it et al.}, Phys.\ Rev.\  C 59, 1663 (1999).
%\cite{Andronic:2009gj}
\bibitem{Andronic:2009gj} A.~Andronic {\it et al.}, Nucl.\ Phys.\  A \textbf{837} 65 (2010). %[arXiv:0911.4806 [hep-ph]].
  %``Hadron Production in Ultra-relativistic Nuclear Collisions: Quarkyonic
  %Matter and a Triple Point in the Phase Diagram of QCD,''
%\cite{Koch:2003pj}
\bibitem{Koch:2003pj}
  V.~Koch,
  %``Strangeness at SIS energies,''
  J.\ Phys.\ G \textbf{30}, S41 (2004).
  %[arXiv:nucl-th/0306037].
  %%CITATION = JPHGB,G30,S41;%%
%\cite{Wheaton:2004qb}
\bibitem{Wheaton:2004qb}
  S.~Wheaton and J.~Cleymans,
  %``THERMUS: A thermal model package for ROOT,''
  Comput.\ Phys.\ Commun.\ \textbf{180}, 84 (2009).
  %[arXiv:hep-ph/0407174].
  %%CITATION = CPHCB,180,84;%%
%\cite{Kraus:2007hf}
\bibitem{Kraus:2007hf}
  I.~Kraus, J.~Cleymans, H.~Oeschler, K.~Redlich and S.~Wheaton,
  %``Chemical Equilibrium in Collisions of Small Systems,''
  Phys.\ Rev.\  C \textbf{76}, 064903 (2007).
  %[arXiv:0707.3879 [hep-ph]].
  %%CITATION = PHRVA,C76,064903;%%
 %\cite{Andronic:2005yp}
\bibitem{Andronic:2005yp}
  A.~Andronic, P.~Braun-Munzinger and J.~Stachel,
  %``Hadron production in central nucleus nucleus collisions at chemical
  %freeze-out,''
  Nucl.\ Phys.\  A \textbf{772}, 167 (2006).
  %[arXiv:nucl-th/0511071].
  %%CITATION = NUPHA,A772,167;%%
%\cite{Maeda:2007cy}
\bibitem{Lopez:2007ms}
  X.~Lopez {\it et al.}  (FOPI Collaboration),
  %``Subthreshold production of $\Sigma$(1385) baryons in Al+Al collisions at
  %1.9$A$ GeV,''
  Phys.\ Rev.\ C \textbf{76}, 052203 (2007).
  %arXiv:0710.5007 [nucl-ex].
  %%CITATION = ARXIV:0710.5007;%%
\bibitem{Maeda:2007cy}
 Y.~Maeda {\it et al.}  (ANKE Collaboration),
  %`Kaon Pair Production in Proton--Proton Collisions,''
 Phys.\ Rev.\  C \textbf{77}, 015204 (2008).
  %%CITATION = PHRVA,C77,015204;%%

%------------------Cascades------------------
%-RHIC
%\cite{Adams:2006ke}
\bibitem{Adams:2006ke}
  J.~Adams {\it et al.}  (STAR Collaboration),
  %``Scaling Properties of Hyperon Production in Au+Au Collisions at
  %sqrt(s_NN) = 200 GeV,''
  Phys.\ Rev.\ Lett.\  \textbf{98}, 062301 (2007).
  %[arXiv:nucl-ex/0606014].
  %%CITATION = PRLTA,98,062301;%%
%--------------------SPS---------------------------
%\cite{Antinori:2004ee}
\bibitem{Antinori:2004ee}
  F.~Antinori {\it et al.}  (NA57 Collaboration),
  %``Energy dependence of hyperon production in nucleus nucleus collisions  at
  %SPS,''
  Phys.\ Lett.\  B \textbf{595}, 68 (2004).
  %[arXiv:nucl-ex/0403022].
  %%CITATION = PHLTA,B595,68;%%
%-AGS--------------------
%\cite{Chung:2003zr}
\bibitem{Chung:2003zr}
  P.~Chung {\it et al.}  (E895 Collaboration),
  %``Near-threshold production of the multi-strange Xi- hyperon,''
  Phys.\ Rev.\ Lett.\  \textbf{91}, 202301 (2003).
  %[arXiv:nucl-ex/0302021].
  %%CITATION = PRLTA,91,202301;%%
%\cite{Kolomeitsev:2009yn}
\bibitem{Kolomeitsev:2009yn}
  E.~E.~Kolomeitsev and B.~Tomasik,
  %``Catalytic phi meson production in heavy-ion collisions,''
  J.\ Phys.\ G \textbf{36}, 095104 (2009).
  %[arXiv:0903.4322 [nucl-th]].
  %%CITATION = JPHGB,G36,095104;%%

%\cite{Schade:2009gg}
\bibitem{Bormio}
M.~Lorenz {\it et al.} (HADES Collaboration),
%Contributed to 48th International Winter Meeting on Nuclear Physics, Bormio, Italy, 25-29 Jan 2010.
PoS BORMIO2010 (2010) 038.
%\bibitem{reisdorf}
% W. Reisdorf {\it et al.} (FOPI Collaboration), Nucl. Phys. A \textbf{848}, 366 (2010).
%\bibitem{geant} GEANT 3.21, Detector Description and Simulation Tool, http://consult.cern.ch/writeup/geant/ (1993).
\end{thebibliography}
\end{document}